\def\etal{{\it et al.\/}}

\def\ie{{\it i.e.\/}}

\def\Pac{{Paczy\'{n}ski\/}}
\def\Mesz{{M\'esz\'aros\/}}

\documentstyle{mn}

\begin{document}

\title[Coronal GRBs and UHECRs]{Coronal Gamma Ray Bursts as the sources of
Ultra High Energy Cosmic Rays?}
\author[M. Vietri]{Mario Vietri $^1$\thanks{E-mail:
vietri@coma.mporzio.astro.it} \\ $^1$
Osservatorio Astronomico di Roma, 00040 Monte Porzio Catone (Roma),
Italy}
\maketitle

\begin{abstract}
I consider the possibility that Ultra High Energy Cosmic Rays are accelerated
in Gamma Ray Bursts located in the Galactic corona, thus circumventing the
problem raised by Greisen--Zatsepin--Kuz'min cutoff.
The acceleration of UHECRs could occur in the pulsars which,
in the coronal GRB model, produce them: the same parameters
that permit fitting GRBs' observations in the model of Podsiadlowski, Rees
and Ruderman (1995) lead to an estimate of the highest achievable energies
corresponding to that of the Bird \etal\/ (1994) event, and to very low
luminosities in cosmic rays. I show that, if the observations of
Milgrom and Usov (1995a) are confirmed, the extragalactic GRBs' model
for the acceleration of UHECRs is untenable, but the same constraint
does not apply to the coronal model. Also, I show that the efficiency
of particle acceleration needs be much smaller (and less demanding) than in
cosmological models of GRBs. Uncertainties remain about the ensuing cosmic
ray spectral distribution. I also briefly discuss observational strategies
to distinguish between the two possibilities.
\end{abstract}

\section{Introduction}

The recent discovery of the two highest energy CRs ever (Bird {\it et al.},
1994, Yoshida {\it et al.}, 1995) has produced a renewal of interest in
the acceleration of Ultra High Energy Cosmic Rays (UHECRs). This is due
to the well--known Greisen--Zatsepin-Kuz'min effect: because of photopion
losses, a proton with $E = 10^{21} \; eV$ at the source needs only cross
a distance $\approx 30 \; Mpc$ (Protheroe and Johnson 1995) to be
slowed down to $E = 3\times 10^{20}\; eV$. Thus it is necessary to identify
acceleration sites which are reasonably close to the Earth.

It has been recently pointed out (Milgrom and Usov 1995a, Waxman 1995a, Vietri
1995) that UHECRs may be accelerated in cosmological Gamma Ray Bursts, GRBs. It
was shown (Vietri 1995) that CRs'energies as high as the largest ever observed
($E \approx 3\times 10^{20} \; eV$, Bird \etal, 1994) can be achieved
in the very short durations ($\approx 1 \; sec$) of GRBs by means of
very efficient first--order Fermi--Bell acceleration in just {\it two} cycles.
However, as it is well--known (Lamb 1995, \Pac\/ 1995),
the nature and location of GRBs is in dispute, the most likely alternative
to the extragalactic model being a distribution of neutron stars
in an extended Galactic corona, at Galactocentric distances $\approx 100\;
kpc$,
and it may be interesting to consider whether UHECRs can be accelerated in
coronal GRBs.

This idea seems to subvert the traditional
view (Cocconi 1956) that UHECRs are of extragalactic origin, the main
evidence for this lying in the change of chemical composition and spectral
slope at the `ankle' ($E = 3\times 10^{18} \; eV$). However, the objects
postulated to give rise to coronal GRBs form a unique population, not just
because their flux and angular
distributions are sharply at odds with those of all known  Galactic
populations and identical to those of all known extragalactic sources,
but also because their Galactocentric distances exceed by an order of magnitude
those of known Galactic objects. Thus, the hiatus that separates Galactic and
extragalactic UHECRs' sources exists in this model as well.
The classical way to try to establish the extragalactic
nature of the sources giving rise to UHECRs
has been to seek the GZK--cutoff in the CRs' flux observed at the Earth.
At present, however, the evidence is lacking statistical significance
(Sigl {\it et al.}, 1995), and does not rule out the coronal model.

It seems thus worthwhile to consider whether UHECRs can be originated in these
nearer models for GRBs. I shall point out below three reasons why this seems
attractive. A discussion will follow in Section 5.

\section{The acceleration of UHECRs in coronal GRBs}

The essential feature of the extragalactic GRB model that is preserved
by the Coronal model is that GRBs are located inside the
Greisen--Zatsepin--Kuz'min (GZK from now on) sphere. The traditional
acceleration sites for UHECRs, Fanaroff--Riley Class II radio galaxies
(Rachen and Biermann 1993) certainly provide an energetically attractive
source of UHECRs, despite the large uncertainties in the estimates of the
proton flux at the source. However, the nearest such galaxy, Pictoris A,
lies at $d_{PA} = 100\; Mpc$. The GZK--radius for $E\ga 2\times 10^{20}\; eV$
is
$R_{GZK} = 20 \; Mpc$ (Protheroe and Johnson 1995), so that the total flux
emitted by Pictoris A is damped by the factor $\exp(d_{PA}/R_{GZK}) \approx
100$. It will be shown later that 1 extragalactic GRB is expected within a
GZK--sphere every $10^3$ years. If this emits equal amounts of energy in
$\gamma$--ray photons and CRs (not just UHECRs), then the expected ratio
of fluxes at the Earth is
\begin{equation}
\frac{f_{GRB}}{f_{PicA}} \approx 0.1 \exp(d_{PA}/R_{GZK}) \approx 10 \;.
\end{equation}
The super--GZK flux by FRII galaxies is even more strongly dominated by
coronal GRBs: all coronal GRBs are located inside $R_{GZK}$, as opposed to just
a fraction $(R_{GZK} H_0 /c)^3 \approx 10^{-6}$, giving a flux of
super--GZK CRs higher by a factor $c/H_0 R_{GZK} \approx 10^2$. Also,
coronal GRBs, exactly like their extragalactic counterparts, are largely
super--Eddington, and hyperrelativistic phenomena like beaming
naturally arise around them, thus mirroring the discussion in Vietri (1995)
that made GRBs attractive as potential sources of UHECRs.

The mechanism for the acceleration of UHECRs in cosmological
GRBs (Vietri 1995) surely does not work for the coronal model, because
the model of an extragalactic GRB as due to the prompt release of a shock's
whole energy immediately after formation, by self--synchro--Compton of
relativistic electrons (\Mesz\/ and Rees 1994)
requires high ISM densities to achieve sufficient efficiencies, $n \approx
1\; cm^{-3}$ (see also
Begelman, \Mesz\/, Rees 1993). Such high baryonic densities
are inconceivable at the distances ($\approx 100 \; kpc$, Podsiadlowski, Rees
and Ruderman 1995, PRR from now on) currently
postulated for the coronal scenario, where the dark matter density is
$\rho_{dm} \approx 3\times 10^{-3} m_H \; cm^{-3}$, and thus most likely
$n \la 10^{-4} \; cm^{-3}$. Models for coronal GRBs cannot simply be those
concocted when GRBs were thought to lie at $\approx 1\; kpc$ from us, because
the greater distance scale implies release of $\approx 10^4$ times more energy.
Recently, however, PRR have shown that a reasonable GRB--generation mechanism
can be identified in the stress--release episodes of the crustal magnetic
field,
provided $B \approx 10^{15} \; G$. This suggests that cosmic rays may be
accelerated in pulsars' magnetospheres. Sigl, Schramm and Bhattacharjee (1994)
give as the highest cosmic ray energy from a pulsar
\begin{equation}
E_{max} = 2\times 10^{20} \; eV \left(\frac{B}{10^{15}\; G}\right) \;.
\end{equation}
The above formula was discussed in this context also by Milgrom and Usov
(1995a); the novel point is the `coincidence' that the same magnetic field is
necessary to explain both the GRB and the UHECR phenomena. While no more
specific predictions can be made because of the lack of a detailed coronal GRB
model, I find this coincidence encouraging.

\section{Coincidences between UHECRs and GRBs}

Another reason why the coronal gamma--ray--burst hypothesis is attractive
comes about when we consider the implications of the work by Milgrom and Usov
(1995a). They discovered that the two super--GZK events (Bird \etal 1994,
Yoshida \etal, 1995) were positionally coincident, within their largish error
boxes, with two strong GRBs which preceded them by $\la 1 \; yr$; on the
basis on this association they proposed that UHECRs are generated in GRBs.
If one believes in this association, the following argument shows that GRBs
are unlikely to be extragalactic. Let us compute the expected rate of GRBs
resulting
in super--GZK events at the Earth. Given that the rate of GRBs is (\Pac, 1993)
$\dot{n} = 30 \; Gpc^{-3} \; yr^{-1}$, and that a Greisen--Zatsepin--Kuz'min
sphere with radius $R_{GZK} = 20 \; Mpc$, for $E=2-3\times 10^{20}\; eV$,
(Protheroe and Johnson 1995) has volume $V_{GZK} = 3\times 10^4 \; Mpc^3$,
I find that a GZK--sphere has a rate of $1$ GRB every $10^3 \;
yr$. However, in the $\approx 10 \; yr$ of combined observational
time of the Fly's Eye and AGASA experiments, two such events have been
already observed. This occurs, in the above model, with probability
$P_2 = (10 q/10^3)^2 \approx 10^{-5}$, where $q$ is the averaged fraction of
sky coverage, which has been taken as $q \la 0.3$. Stated another way,
for these low values of $R_{GZK}$ and of the time--delay, most of the
time (always but just once in $10^3 \; yr$)
we should not be able to observe any super--GZK CR. It follows that, if the
proposed observational connection and time delay ($\la 1\; yr$) were to
be confirmed, the sources that produce UHECRs could not be extragalactic
GRBs.

The above observation is of course unlikely even when viewed from the other
angle, that of GRBs. Milgrom and Usov (1995a) had at their disposal $\approx
3\; yr$ of the BATSE catalog. This implies then that the probability of
seeing $2$ GRBs located within $R_{GZK}$ is also given by
$(3/10^3)^2 = 10^{-5}$, fortuitously equal to $P_2$ above.
Another way to look at the same problem is to compute the total energy released
by the two GRBs that Milgrom and Usov (1995a) associated with the two
highest energy CRs. They have fluences of $4\times 10^{-5} \; erg \; cm^{-2}$
and $3\times 10^{-4} \; erg \; cm^{-2}$, which assuming a distance $=
R_{GZK} = 20 \; Mpc$, correspond to a total energy release of $10^{48} -
10^{49} \; erg$. This is low when compared with the average
energy released by GRBs, $4\times 10^{51} \; erg$ (Piran 1992). While
it is certainly possible that the GRBs' luminosity function is broad,
still the smallness of the total energy released computed thusly is in keeping
with the argument developed above.

It should be noticed that the previous argument is based upon the smallness
of $R_{GZK} = 20 \; Mpc$. Protheroe and Johnson (1995)
compare their results with
several previous computations, and it is apparent from their Fig. 4 that
theirs is the largest value in the energy range of interest here, some
authors having obtained values as low as $R_{GZK} = 9 \; Mpc$.

Independently of the actual time--delay, it seems very unlikely to me that
any GRB in the BATSE catalog can come from within the GZK--sphere. One way out
of this predicament is, of course, if the time--delay is actually much longer
than proposed by Milgrom and Usov (1995a); in this case there need be no
GRB from inside the GZK sphere in the BATSE catalog, and, if it is assumed that
CRs from the same source are spread out over a time comparable to the
time--delay, then perhaps a hundred distinct extragalactic
GRBs as sources of UHECRs should be seen at all times (see later on).
More data should settle this issue.
In the following I shall assume the connection and the time--delay
$\approx 1\; yr$ to be correct. In this
case the above predicament would be relieved if coronal GRBs were to
produce super--GZK CRs, first because photopion and photoelectron losses
within the Galaxy are entirely negligible (and thus several GRBs from
within the UHECR error box can be the
putative fathers of the super--GZK events), and second because there is no
need that the objects giving rise to GRBs produce UHECRs only during
fireballs: they could instead produce a steady flux of CRs.
Under the coronal hypothesis, the time--delays of $\approx 1 \; yr$ can easily
be accommodated (Milgrom and Usov 1995a).

\section{The efficiency}

Another reason why a Coronal GRB origin for UHECRs is attractive is due
to the fact that it was quickly realized (Vietri 1995,
Waxman 1995a, Milgrom and Usov 1995b) that accounting for all UHECRs observed
at Earth requires that each extragalctic GRB releases approximately equal
amounts of energy
in $\gamma$--ray photons and in UHECRs, \ie, very high efficiency.
In order to see how serious this efficiency problem is, it is convenient to
compare the expected production by GRBs of UHECRs in the range
$10^{19} \; eV < E < 10^{20} \; eV$
with that deduced from observations (Waxman 1995b),
$\epsilon_{obs} = 5\times 10^{-37} \; erg \; s^{-1} \; cm^{-3}$. The observed
rate of GRBs is $\dot{n} = 30 \; Gpc^{-3} \; yr^{-1}$ (\Pac, 1993). The average
GRB total energy output in photons $E_{\gamma}$
is $4\times 10^{51} \; erg$ (Piran 1992). This
yields the total energy release rate in $\gamma$--ray photons. To estimate
the release rate of energy in UHECRs I proceed as follows.
The fraction $q$ of the directed kinetic
energy that shocks convert into CRs, compared to that converted into
thermal motions (which in GRBs is promptly dissipated into $\gamma$--ray
photons) is very uncertain, and is
variously estimated (Draine and McKee 1993) between $0.03$ and $0.5$.
I shall take the larger value. The spectrum emitted by a
cosmological source is derived from observations (Waxman 1995b) as $\propto
E^{-2.3}$. I shall consider a harder spectrum, $\propto E^{-2}$, so as to
continue to overestimate the production of UHECRs by GRBs. Then the total
fraction of energy released in cosmic rays, and channeled in the
range discussed by Waxman, $10^{19} \; eV < E < 10^{20} \; eV$, is $p =
\ln 10 / \ln (E_{max}/E_{min})$. I shall conservatively take $E_{max} =
10^{20}\; eV$, and $E_{min} = 10^{15} \; eV$. This last value comes from
this argument. Vietri (1995)
took as would--be CRs the extreme Boltzmann tail
of the just--shocked ISM protons, for which $\gamma \approx
10^2 - 10^3$ in the shell frame. In the lab frame these protons would
appear as CRs with $E \approx 10^{13} - 10^{15} \; eV$, corresponding to the
range in $\gamma$. Together, these two factors imply
a relative efficiency of UHECRs to $\gamma$--ray photons of $\eta =
E_{UHECR}/E_{\gamma} = q p \approx
1/10$. The comparison with observations yields
\begin{equation}
\frac{\epsilon_{GRB}}{\epsilon_{obs}} =
\frac{\dot{n} E_{\gamma} \eta}{\epsilon_{obs}} =
0.03 \;,
\end{equation}
despite my attempts at maximizing the contribution of GRBs. Thus it seems
likely that, in order to reproduce observations,
GRBs must overproduce UHECRs with respect to conventional models,
by a factor $10-100$, {\it i.e.\/}, $\eta \approx 1-10$.
It should be emphasized that large efficiencies are not impossible:
the above estimate is very uncertain, and it refers to steady--state, newtonian
shocks because no analogous computations are known to me in the
time--dependent,
relativistic regime. However, in the face of the daunting task of raising
the efficiency by about two orders of magnitude, it seems worthwhile to
consider the alternative hypothesis that GRBs originate in the Galactic corona.

The efficiency requirement becomes immediately less stringent in the
coronal model. In fact, if UHECRs are generated by cosmological GRBs,
only those originating within $R_{GZK}$, the Greizen--Zatsepin--Kuz'min
radius,  of the Milky Way can reach us.  This is because UHECRs with $E \ga$
a few times $10^ {19}\; eV$ loose energy by photopion and photoelectron
production off
CMBR photons. These nearby GRBs account for a fraction $f \approx R_{GZK}/
(c/H_0)$ of the whole flux at Earth. On the other hand, if GRBs are
located in the Galactic Corona, all GRBs generate UHECRs that reach the Earth.
The coronal GRBs' energy release ($E_{GRB} \approx 10^ {41}\; erg$) is
determined so that the flux of photons at Earth is equal to that
in the cosmological model; thus the coronal model with the same efficiency
factor $\eta$ as the extragalactic model produces  an
UHECRs' flux at Earth higher by the factor $1/f$ with respect to the
extragalactic model. With $R_{GZK} = 20 \; Mpc$ (Protheroe and Johnson 1995),
for $E \approx 3\times 10^{20} \; eV$, the overproduction is $1/f \approx
10^2$. This implies that, in the coronal model, the fitting of the observed
UHECRs' flux can be achieved with an  efficiency $\eta$ reduced by the factor
$f$ to $\eta \approx 0.03$.

\section{Discussion}

The low efficiency $\eta \approx 0.03$ discussed above is rather rewarding in
the case in which UHECRs are generated during GRBs. However, another, more
effective way for the low value of $f$ to ease our luminosity quandaries
occurs if the UHECR--production is steady, within the PRR model. In fact, PRR
postulate that every pulsar remains active for up to $10^{10}\; yr$, producing
$\approx 10^6$ stress--release episodes within its lifetime. This means a GRB
every $10^4\; yr$, which equals a time--averaged GRB--luminosity of $\approx
3\times 10^{29} \; erg\; s^{-1}$. With $\eta \approx 0.03$, this leads to a
continuous UHECR--luminosity of $L_{UHECR} \approx 10^{28}\; erg \;
s^{-1}$.

However, let me remark that, if the release of the UHECRs were coincident
with the event leading to the GRB, the acceleration of UHECRs could use as
an energy source the very same one of the GRB, and it would be energetically
insignificant since $\eta \approx 0.03$. This would also leave room for the
acceleration of several other decades of CRs' energy beyond the one
considered, $10^{19} \; eV < E <10^{20} \; eV$, even assuming a softer
spectrum. It is more difficult to identify an energy source in the case of
continuous acceleration of CRs: with a field $B \approx 10^{15} \; G$, the
rotational kinetic energy of the pulsar must have been exhausted very early
indeed.

Another observational feature, the dominance of protons with respect to
heavier nuclei beyond the ankle ( $E=3\times 10^{18}\; eV$,
Bird \etal, 1994, 1995), which is often cited as evidence for the extragalactic
origin of UHECRs, can be accommodated easily  within this model. Iron
nuclei with $E \approx 10^{20}\; eV$ have gyroradii $r_L \approx 5 \; kpc$
in the Galactic magnetic field, so that they are essentially confined
around the pulsars producing them. All diffusion processes tend to push them
away from us, and into the IGM, thus making them unobservable to us.
Protons of comparable energy have much larger gyroradii ($\approx 100\; kpc$),
which allow them to penetrate into the inner Galaxy.

The most significant weakness of the coronal model is that the expected
spectrum is not necessarily
close to $E^{-2}$, as it automatically is every time CR
acceleration at shocks is invoked. This point is clearly in need of
further investigation.

Neglecting the thornier question of establishing a connection between
UHECRs and GRBs, I discuss now a comparison between the extragalactic and
coronal models for GRBs as sources of UHECRs. As stated earlier, spectral
evidence can distinguish between the two models, and, with the
arrival of the next generation of detectors, the test is feasible
(Sigl {\it et al.\/}, 1995). Another significant observational difference
between the two models occurs when we consider the angular distribution of
super--GZK events. In the coronal model, they must be isotropic because the
GRB model (PRR) is designed to fit the observations (Meegan {\it et al.},
1992),
except for the very small dipole anisotropies predicted, $\approx$ a few
times $10^{-2}$. In the extragalctic model, super--GZK CRs must occur within
a GZK sphere, with $R_{GZK} \la 30 \; Mpc$. Within this distance the peculiar
velocity of the Galaxy is formed (Scaramella, Vettolani and Zamorani 1994), so
that we expect larger anisotropies. Here, I cannot help but notice the irony
that it is the detection of an isotropy that would favor the local model,
the only such case known to me in astronomy.

For sufficiently large detectors,
the total number of independent directions of arrival in the coronal model
must be at least as large as that of known GRBs, a few thousands. This
differs sharply from the number expected for the extragalactic model. From
within a sphere $R \la 30\; Mpc$ (corresponding to the path--length of a CR
which started out with $E = 10^{21} \; eV$ and is observed with $E = 3\times
10^{20} \; eV$) we expect in fact $\approx$ 1 GRB every 300 years. The most
reasonable time--delay over $R_{GZK}$ is given by
\begin{equation}
\bigtriangleup\!t = 3\times 10^4 \; yr \;
\left(\frac{R}{30\; Mpc}\right)^3
\left(\frac{B}{1\; nG}\right)^2
\end{equation}
and this, assuming that the UHECRs coming from a GRB are spread over a
timescale $\approx \bigtriangleup\!t$, means that $\approx 100$ independent
sites from which super--GZK CRs are coming, are visible at all times. Thus
about an order of magnitude separates the number of different acceleration
sites visible in the two models, the coronal one being the more densely
populated.

Assuming the validity of the extragalactic model, I would like to point out
that
the above--determined rate of GRBs inside the GZK--sphere, $\dot{n}_{GZK}
= 1/10^3\; yr$, allows a measurement
of the time--delay between GRBs and UHECRs, in a different range than the one
($\approx 1 yr$) discussed by Milgrom and Usov (1995a). In fact, suppose that
experiments with perfect sky coverage were to reveal the existence of $N$
small regions on the plane of the sky from which super--GZK events seem
to arrive. Then, we can use the inverse of the rate above, $1/\dot{n}_{GZK}$,
as a universal clock to state that the time delay is $\approx N/\dot{n}_{GZK}$.
The areas of the small regions is limited from below by instrumental
resolution,
or by the deflections of particles along their flight--path to our detectors.
Since, from conventional estimates of the magnetic field it is found that
deflection angles of a few degrees are most likely for UHECRs (Sigl, Schramm
and Battacharjee 1994), we expect to be able to resolve at most $\approx 10^3$
such clusters. This means that the time--delay that can be measured is in
the range $1/\dot{n}_{GZK} - 10^3/\dot{n}_{GZK}$, \ie, $10^3-10^6 \; yr$.
A detailed study of this measurement will be presented elsewhere.

In summary, the model of PRR, designed to fit the
observations of GRBs, also naturally accounts for the energy of the super--GZK
Crs observed so far. The same model also avoids the efficiency problem
of the extragalactic competitor, and is consistent with the statistical
significance of the coincidences found by
Milgrom and Usov (1995a). Angular distribution properties are sufficiently
distinct from those of the extragalctic model to make discrimination of the
two models feasible with the next generation of detectors.

\end{document}